\documentclass[superscriptaddress,showpacs,pre,twocolumn]{revtex4-1}
\usepackage{subfigure}
\usepackage{graphicx,graphics,bm,mathrsfs,color}% Include figure files
\usepackage{dcolumn}% Align table columns on decimal point
\usepackage{bm}% bold math
\usepackage{hyperref}% add hypertext capabilities
\usepackage[mathlines]{lineno}% Enable numbering of text and display math
\usepackage[utf8]{inputenc}

\begin{document}

\preprint{APS/123-QED}

\title{Particle creation in FRW with variable $q$,  $G$ and $\Lambda$}

\author{Özgür Sevinc$^{1,2}$ and Ekrem Aydiner} \email{ekrem.aydiner@istanbul.edu.tr}
\affiliation{Department of Physics, Istanbul University, TR34134 Istanbul, Turkey\\
$^{2}$Department of Electric and Electronic, Istanbul Yeni Yüzyıl University, TR34010 Istanbul, Turkey}

\date{\today}% It is always \today, today,

\begin{abstract} %%%%%%%%%%%% Aydiner Date: Dec 15, 2016 h:09:25 istanbul 
In this study, the mechanism of particle creation using varying gravitational and cosmological constants and deceleration parameter has been studied for Friedman-Robertson-Walker at high dimensions to explain early deceleration and present accelerating phases. In order to investigate the dynamics of two phases, we have considered two different ansatz for the scale factor of the form  $a(t)=\sqrt{t^{\alpha}e^{t}}$ and $a(t)=\sqrt{\sin h(kt)}$ which are general form of power law expansions. Firstly we modified $d$-dimensional field equations depend on time introduce general formulation of particle creation and entropy generation mechanisms. We investigate time dependence of the several cosmological constant and quantities such as particle creation $\psi$ and entropy $S$, gravitational constant $G$, cosmological term $\Lambda$, energy density $\rho$, deceleration parameter $q$ etc. It is shown that all constant and other quantities, except the cosmological constant $G$ and entropy $S$, characteristically decrease with time for two scale factors in all dimensions. However, the cosmological constant $G$ and entropy $S$ increase with time. Additionally, it is shown that the cosmological constant $\Lambda$ is unexpectedly independent of particle creation mechanism.

\end{abstract}

\maketitle

%%%%%%%%%%%% Aydiner Date: May 14, 2017 h:11:15
\section{Introduction}

Einstein saw that the universe could not be stationary because of the solution of the field equations, and he created a stationary universe model in the frame of the homogeneous and isotropic RW metric by adding the cosmological constant ($\Lambda$) which can act as a driving force to the field equations. After Hubble announced that the radial shift values of sky island radiation were directly proportional to distance, Einstein's stationary universe model was replaced by an expandable universe model based on the RW metric that Friedman’s work in 1922 and 1924 revealed. This model, known today as FRW models, forms the basis of the Standard Big Bang universe model. The success of this model comes from the Hubble expansion law, light element abundances observed in the environment and cosmic microwave background (CMB) radiation. However, it has been noticed that this model has some important problems. These problems can be solved by passing a stage similar to the exponentially expanding universe model that Sitter obtained using $\Lambda$ in two of his studies. This laid the foundations for inflation models that were obtained by using energy sources that acted as $\Lambda$ in the early periods of the universe. Today's cosmological data shows that the early universe passed through an inflationary period, then evolved in accordance with Standard models, and the present universe accelerated by the influence of a dark energy acting as $\Lambda$ and has an almost spatially flat geometry. Observations are able to give the proportions of the physical components of the universe with great sensitivity \cite{Bergman1968,Linde1974,Perlmuntter1999,Abazajian2004,Peebles2003}. However, the cosmological constant is assumed to be very small. To solve this problem, it can be assumed that $\Lambda$ is not constant but decreases continuously over time. Recent observations and cosmological indications in particle physics show that $\Lambda$ may be more dynamic than constant. The effect of a variable $\Lambda$ on the scale of the scale factor in the FRW universe is investigated. There are a number of studies that cosmological constant has been reviewed as a function of time \cite{Vishwakarna2002,Beesham1999,Peebles1988}.

The fact that the gravity constant $G$ is decreasing has yielded quite interesting results in many theories.  
These studies, starting with Dirac, have gained speed nowadays \cite{Dirac1937,Brans1961}. Prigogine and his colleagues were the first people who addressed the particle creation problem \cite{Prigogine1988,Prigogine1989}. As a result, today, this idea is very popular and there are many available studies in the literature. Particle creation occurs with the emerging of an additional pressure in the energy-momentum tensor. In the frame of FRW, the general structure of the first law of thermodynamics is used for open systems defined as energy flows from gravitational fields to material fields. This is considered as an additional negative pressure, and the energy-momentum tensor has been reinterpreted. According to this model, the reason for the expanding space-time during the reversible processes is the growth of entropy by producing matter. Indeed, the negative pressure of the creation of particles can play the role of dark energy, which explains the acceleration of the universe. 

So far, based on this idea, universe dynamics has been investigated for a few scale factors \cite{Singh2012,Calvao1992,Alcaniz1999,Lima2008,Steigman2009}. However, in order to investigate early deceleration and present accelerating phases of the universe depending on particle creation, more realistic scale factors such as $a(t)=\sqrt{t^{\alpha}e^{t}}$ and $a(t)=\sqrt{\sin h(kt)}$ have never used. These scale factors are commonly used to explain early deceleration and present accelerating phases in the literature. Therefore, in this study, in order to see the effect of these scale factors on the evolution of the universe and we studied particle creation and entropy generation in higher dimensional FRW model with varying gravitational constant $G$ and cosmological term $\Lambda$ for these scale factors. Additionally, we investigate time dependence of the several cosmological quantities such as particle creation $\psi$ and entropy $S$, energy density $\rho$, deceleration parameter $q$ etc. We see that all constant and other quantities, except the cosmological constant $G$ and entropy $S$, characteristically decrease with time for two scale factors in all dimensions. However, cosmological constant $\Lambda$ is independent of particle creation mechanism.

The paper is organized as follow: In Section 2, we discuss modified field equations, In Section 3, we present main formalism of the particles creation mechanism by using modified field equations. In Section 4, we analyzed the several cosmological constant and quantities such as particle creation $\psi$ and entropy $S$, gravitational constant $G$, cosmological term $\Lambda$, energy density $\rho$, deceleration parameter $q$ by using numerical solutions and present all results. Finally we discussed and conclude the obtained results in Section 5.

\section{Model and Fields Equations}

We would like to consider $(m + 2)$ dimensional homogeneous, isotropic and smooth space-time structures represented by Friedman-Robertson-Walker (FRW);
\begin{eqnarray} \label{FE-ds}
ds^{2} = dt^{2}-a^{2}(t) \left[dr^{2} + r^{2} dx^{2}_{m}  \right] 
\end{eqnarray}
where $a(t)$ is scale factor, and $dx^{2}_{m}$;
\begin{eqnarray} \label{FE-dx}
dx^{2}_{m} = d\theta^{2}_{1} + \sin^{2}\theta_{1} d\theta^{2}_{2}+...+ sin^{2}\theta_{m-1} d\theta^{2}_{m}
\end{eqnarray} 
Assuming that the universe is filled with perfect fluid, the energy-momentum tensor is expressed;
\begin{eqnarray} \label{FE-Tij}
T_{ij} = (\rho + p) u_{i} u_{j} - p g_{ij}
\end{eqnarray}
where $\rho$ is the energy density, $p$ is cosmic fluid pressure and $u_{i}$ is the four-velocity vector which has $u_{i}u^{j}=1$.

Time-independent Einstein field equations with gravitational constant $G$ and cosmological constant $\Lambda$ are;
\begin{eqnarray} \label{FE-Rij}
R_{ij}-\frac{1}{2}g_{ij}R = -8 \pi G T_{ij} - \Lambda g_{ij}
\end{eqnarray}
where $R_{ij}$, $g_{ij}$ and $R$ are the Ricci, the metric tensor and the Ricci scalar, respectively. Using the Eqs.~(\ref{FE-ds}) and (\ref{FE-Tij}), we get time-dependent i.e., modified Einstein field equations as in the following:
\begin{eqnarray} \label{FE-x5}
\frac{m(m+1)}{2}\left(\frac{\dot{a}}{a}\right)^{2} = 8 \pi G(t) \rho + \Lambda(t)
\end{eqnarray}
and 
\begin{eqnarray} \label{FE-x6}
m \frac{\ddot{a}}{a} + \frac{m(m-1)}{2}\left(\frac{\dot{a}}{a}\right)^{2} = - 8 \pi G(t) p + \Lambda(t)
\end{eqnarray}
where $dot$ shows derivation according to cosmic time. When derivation of the Eq.\,~(\ref{FE-x5}) is taken,
\begin{eqnarray} \label{FE-x7}
\frac{m(m+1)}{2} \left[ \frac{2\dot{a}}{a} \left(\frac{\ddot{a}}{a} - \frac{\dot{a}^{2}}{a^{2}}\right)  \right] = 8 \pi \left(\dot{G} \rho + G \dot{\rho} \right) + \dot{\Lambda}
\end{eqnarray}
is obtained. When we sum up the Eqs.~(\ref{FE-x5}) and (\ref{FE-x6}) and multiply the Eq.~(\ref{FE-x5}) by $(-1)$, 
\begin{eqnarray} \label{FE-x8}
m \left[\frac{\ddot{a}}{a} -\left(\frac{\dot{a}}{a} \right)^{2}  \right] = -8\pi G \left(p + \rho \right) 
\end{eqnarray}
are obtained. When we write the Eq.~(\ref{FE-x8}) to the Eq.~(\ref{FE-x7}), equation of continuity 
\begin{eqnarray} \label{FE-x9}
\dot{\rho} + \left( m+1 \right) \left(p+\rho \right) \frac{\dot{a}}{a} +\rho \frac{\dot{G}}{G} + \frac{\dot{\Lambda}}{8\pi G} = 0 
\end{eqnarray}
is obtained. Also, we use Eqs.\,(\ref{FE-x5}) and (\ref{FE-x9}), and we obtain general form of the energy density and cosmological constant as in the following:
\begin{eqnarray} \label{FE-x10}
p + \rho(t) = - \frac{m\dot{H}}{8\pi G}
\end{eqnarray}
and
\begin{eqnarray} \label{FE-x11}
\Lambda(t) = \frac{m(m+1)}{2}H^{2} - 8 \pi G \rho \ .
\end{eqnarray}
where $H=\frac{\dot{a}}{a}$.

\section{Thermodynamics of the Particle Creation}

In the early period of the universe, we can think of the particle structure, noninteracting particle density number $n$, and the relativistic fluid adapting to the following the equation of state
\begin{eqnarray} \label{N-x12}
p = \omega \rho
\end{eqnarray}
and 
\begin{eqnarray} \label{N-x13}
n = n_{0} \left(\frac{\rho}{\rho_{0}}  \right)^{\frac{1}{1+\omega}}
\end{eqnarray}
where $\omega$ is a parameter of the state equation that takes values in the range of $-1<\omega \le 1$. However, Supernova (SNe) Ia, and Cosmic background radiation (CMB) data are in the range of $-1.3<\omega\le-0.79$ for the accelerating universe. The present values of the particle number and energy density are respectively $n_{o} \geq 0$ and $\rho_{0} >0$. The particle density number provides equilibrium equations \cite{Prigogine1988,Prigogine1989}
\begin{eqnarray} \label{N-x14}
\dot{n} + \left(m +1 \right)m H = \psi (t) n
\end{eqnarray}
where $\psi(t)$ is the particle creation rate with dimension $t^{-1}$ \cite{xHarko1999,xLima1990}. On the other hand, $\psi(t) >0$, $\psi(t)<0$ and $\psi(t)=0$ respectively show particle source, particle disappearance and situations with no particle production. When we rewrite Eqs.\,(\ref{FE-x9}) and (\ref{N-x13}) with respect to Eq.\,(\ref{N-x12}), we obtain respectively as
\begin{eqnarray} \label{N-x15}
\frac{\dot{\rho}}{\rho} = - (m+1) (1+w) H - \left( \frac{\dot{G}}{G} + \frac{\dot{\Lambda}}{8 \pi G \rho}   \right) = 0
\end{eqnarray}
\begin{eqnarray} \label{N-x16}
\frac{\dot{n}}{n} = \frac{1}{1+w} \frac{\dot{\rho}}{\rho} \ .
\end{eqnarray}
Finally, we use Eqs.\,(\ref{N-x14})-(\ref{N-x16}), we get the general form of the particle creation function as
\begin{eqnarray} \label{N-x17}
\psi(t) = -\frac{1}{(1+w)G} \left( \frac{\dot{\Lambda}}{8\pi \rho} + \dot{G} \right) \ .
\end{eqnarray}

The entropy $S$ produced depends on the particle creation at temperature $T$ is given by the following equation
\begin{eqnarray} \label{E-x1}
T \frac{dS}{dt} = \frac{d(\rho V)}{dt} + p \frac{dV}{dt}
\end{eqnarray}
where $V=a^{m+1}$ is the spatial co-moving volume For cosmological fluid which the density and pressure depend on temperature only, the entropy takes the following form
\begin{eqnarray} \label{E-x2}
S = \frac{(\rho + p) a^{m+1}}{T} = \frac{(1+w) \rho a^{m+1}}{T}
\end{eqnarray}
From Eqs.\,(\ref{E-x1}) and \,(\ref{FE-x9}) we get
\begin{eqnarray}\label{E-x3}
T \frac{dS}{dt} = - \frac{1}{G(t)} \left[ \frac{\dot{\Lambda(t)}}{8\pi G \rho} + \dot{G}(t)  \right] \rho(t) a^{m+1}
\end{eqnarray}
or equivalently
\begin{eqnarray}\label{E-x4}
\frac{dS}{dt} = \frac{(1+w) \rho a^{m+1}}{T} \psi(t) \ .
\end{eqnarray}
From Eqs.\,(\ref{E-x3}) and \,(\ref{E-x4}), the entropy as a function of the particle creation rate is given by
\begin{eqnarray}\label{E-x5}
S(t) = S_{0} e^{\int\psi(t) dt}  
\end{eqnarray}
where $S_{0}$ is the initial entropy. In this section, we summarized the mechanism of the particle creation and time-dependent entropy in the FRW framework based on previous studies. The comprehensive discussions can find in Refs.\,\cite{Prigogine1988,Prigogine1989,xHarko1999,xLima1990} and therein references.

\section{Models and Their Cosmological Solutions}

In this section, we consider two different time dependent scale factor, mentioned in the introduction, such as $a(t)=\sqrt{t^{\alpha}e^{t}}$ and $a(t)=\sqrt{\sin h(kt)}$ and we discuss cosmological solutions of particle creation $\psi$ and entropy $S$. Additionally, we discuss the time evolution behavior of the several cosmological constant and quantities such as gravitational constant $G$, cosmological term $\Lambda$, energy density $\rho$, deceleration parameter $q$. Analytical and numerical results are presented below.

\subsection{Model I}

In order to investigate effect of the particle creation on the universe dynamics, we firstly consider the ansatz for the scale factor of the form;
\begin{eqnarray} \label{N-x18}
a(t) = \sqrt{t^{\alpha}e^{t}}
\end{eqnarray}
where $\alpha$ is a positive constant and if $\alpha = 0$ is selected, the Eq.\,(\ref{N-x18}) is reduced to the exponential change law as $a(t)=\sqrt{e^{t}}$. Here we note that the time evolution character of the scale factor can be determined by the deceleration parameter $q$. It is assumed that the universe accelerates for $q$ takes value in the interval $-1<q<0$, expands slowly for $q=0$, and  evaluates with constant speed for $q>0$. One can find a lot of study related to deceleration parameter in the literature \cite{Perlmutter1998,Perlmutter1999,Riess1998,Schuecker1998,Mak2002,Ahmed2014,Akarsu2012,Amirhashchi2011,Pradhan2006,Pradhan2007,Pradhan2014a,Saha2012}. 

The general form of the deceleration parameter $q$ is defined with the $H$ Hubble parameter as
\begin{eqnarray} \label{N-x19}
q=-1-\frac{\dot{H}}{H^{2}} = -\frac{a \ddot{a}}{\dot{a}^{2}} = - \frac{\ddot{a}}{aH^{2}} 
\end{eqnarray}
where $H=\dot{a}/a$. This parameter for scale factor in Eq.~(\ref{N-x18}) can be obtained as
\begin{eqnarray} \label{N-x20}
q=\frac{2\alpha}{(\alpha +t)^{2}}-1  \ .
\end{eqnarray}
It can be seen from Eq.~(\ref{N-x20}) that the parameter $\alpha$ determine the time evolution charecter of the deceleration parameter in Eq.~(\ref{N-x18}). This conditions satisfy  $q>0$; $t<(\sqrt{2\alpha}-\alpha)$ and $q<0$; $t>(\sqrt{2\alpha}-\alpha)$. For $0<\alpha<2$ the 
model evolves from the deceleration phase to the acceleration phase.
The current observations in SNe Ia show that in some regions the deceleration parameter is $-1<q<0$. This analyzing shows that the selected scale factor can be used.

Afterward, we can discuss the other physical quantities we mentioned above. If we take the derivative of the scale factor with respect to time and we define the Hubble parameter $H$;
\begin{eqnarray} \label{N-x21}
H=\frac{\dot{a}}{a}=\left(\frac{\alpha}{2t}+\frac{1}{2}\right) \ .
\end{eqnarray}
Using the Eqs.~(\ref{FE-x10})-(\ref{N-x12}) and (\ref{N-x21}) we obtain the energy density and cosmological constant in order as in the following:
\begin{eqnarray} \label{N-x22}
\rho(t) = \frac{m\alpha}{16\pi G(1+\omega)} \frac{1}{t^{2}}
\end{eqnarray}
\begin{eqnarray} \label{N-x23}
\Lambda(t) = \frac{m(m+1)}{8} \left[ \frac{\alpha}{t} + 1  \right]^{2} - \frac{m \alpha}{2(1+\omega)} \frac{1}{ t^{2}}
\end{eqnarray}
It is clear that $\omega \ne -1$ is required for both solutions to be valid. The energy density and the behavior of the Cosmological constant are decreasing infinitely at $t = 0$ and go to zero for $t \rightarrow \infty$. 

Vacuum energy density $\rho_{\Lambda}$;
\begin{eqnarray} \label{N-x24}
\rho_{\Lambda}(t) = \frac{1}{8\pi G} \Bigg\{ \frac{m(m+1)}{8} \left[ \frac{\alpha}{t} + 1  \right]^{2} - \frac{m \alpha}{2(1+\omega)} \frac{1}{t^{2}} \Bigg\}
\end{eqnarray}
is obtained. Density parameters for Matter and Vacuum are;  
\begin{eqnarray} \label{N-x25}
\Omega_{matt}(t)=\frac{\rho}{\rho_{c}} =\frac{2\alpha}{(1+m)(1+\omega)}\frac{1}{2t^{2}}\left[ \frac{\alpha}{2t} + \frac{1}{2}  \right]^{-2}
\end{eqnarray}
\begin{eqnarray} \label{N-x26}
\Omega_{\Lambda}(t)= 1- \frac{2\alpha}{(1+m)(1+\omega)}\frac{1}{2t^{2}}\left[ \frac{\alpha}{2t} + \frac{1}{2}  \right]^{-2} \ .
\end{eqnarray}
Here $\rho_{c}$ is defined as in the following:
\begin{eqnarray} \label{N-x27}
\rho_{c} = \frac{m(m+1)H^{2}}{16 \pi G}
\end{eqnarray}
Total density parameter which is the sum of the density parameters for Matter and Vacuum is;
\begin{eqnarray} \label{N-x28}
\Omega_{T} = \Omega_{matt} + \Omega_{\Lambda} = 1
\end{eqnarray}
This solution is similar to Einstein field equations and is compatible with inflation scenarios \cite{Singh2012}.
When we use the Eqs.\,(\ref{N-x15}) and (\ref{N-x22}) into Eq.\,(\ref{N-x17}), particle creation function 
\begin{eqnarray} \label{N-x29}
\psi(t) = \frac{1}{2(1+\omega)} \Big\{ -\frac{4}{t} + (1+m) (1+\omega) \left( \frac{\alpha}{t} + 1 \right)  \Big\}
\end{eqnarray}
is obtained. Finally, the entropy production during the particle creation era can be obtained from Eqs.\,(\ref{E-x5}), (\ref{N-x18}) and (\ref{N-x29}) as
\begin{eqnarray} \label{N-e1}
S(t) = S_{0} t^{\frac{\alpha(1+m)(1+w)-4}{2(1+w)}} e^{\frac{(1+m)}{2}t} \ .
\end{eqnarray}

In the case of no particle production, i.e., $\psi(t)=0$, we get the usual particle conservation law of the standard cosmology. For holding conservation law, Eq.\,(\ref{FE-x11}) decouples to the following equation:
\begin{eqnarray} \label{N-x30}
\dot{\rho}+(m+1)(w+1)\rho H = 0
\end{eqnarray}
\begin{eqnarray} \label{N-x31}
\dot{\Lambda}(t) + 8\pi \rho \dot{G}(t) = 0
\end{eqnarray}
Also, Eq.\,(\ref{N-x30}) gives 
\begin{eqnarray} \label{N-x32}
\rho = \rho_{0} a^{-(m+1)(w+1)}
\end{eqnarray}
where $\rho_{0}$ is a positive constant. Using Eq.\,(\ref{N-x18}) into Eq.\,(\ref{N-x32}) we get
\begin{eqnarray} \label{N-x33}
\rho = \rho_{0} (t^{\alpha} e^{t})^{-\frac{(m+1)(1+w)}{2}}
\end{eqnarray}
Using Eq.\,(\ref{FE-x5}) and (\ref{N-x33}), one gets the evolution of the gravitational and cosmological constants, which are given by
\begin{eqnarray} \label{N-x34}
G = \frac{m \alpha}{16 \pi \rho_{0}(1+w)}e^{\frac{t(m+1)(1+w)}{2}}t^{\frac{a (m+1)(1+w)}{2}-2}
\end{eqnarray}
and
\begin{eqnarray} \label{N-x35}
\Lambda = \frac{m(m+1)}{8} \left[ \frac{\alpha}{t} + 1 \right]^{2} - \frac{m \alpha}{2(1+w)} \frac{1}{t^{2}} \ .
\end{eqnarray}
So far, we obtained analytical results for Model I. Now we present numerical solutions.  

The consistency of the Model I is very important in order to achieve changes over time for various cosmological variables. The figures are formed with the appropriate values for the parameters $\alpha$ and $m$, respectively $0<\alpha<2$ and $m>1$.  
\begin{figure} [h!]
	\centering
	\includegraphics[width=6cm]{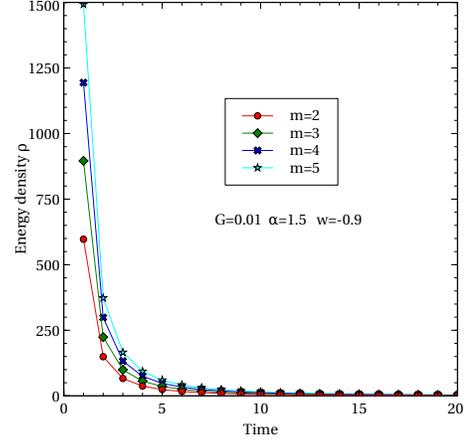}
	\caption{Energy density versus time for $\alpha=1.5$, $w=-0.9$ and $G=0.01$.}
	\label{fig-1}
\end{figure}
\begin{figure} [ht!]
	\centering
	\includegraphics[width=6cm]{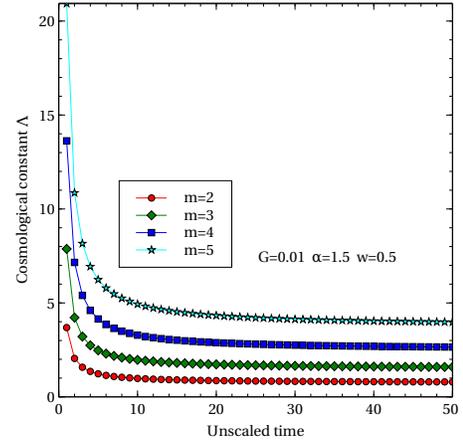}
	\caption{Cosmological constant versus time for $\alpha =1.5$, $w = 0.5$ and $G = 0.01$.}
	\label{fig-2}
\end{figure}

In order to obtain time dependent behavior of the energy density we numerically solve Eq.\,(\ref{N-x22}) for arbitrary $\alpha=1.5$, $w=-0.9$ and $G=0.01$ values. Numerical results for different dimensions ($m=2,3,4,5)$ are given in Fig.\,\ref{fig-1}. It can be seen from Fig.\,\ref{fig-1} that energy density slowly decrease with time and reach to zero. This behavior of the energy density consistent with the prediction of the standard cosmology data. Figure also provide that time evolution character in all dimension are similar.  

To see time evolution of the cosmological constant $\Lambda$, we solve Eq.\,(\ref{N-x23}). The numerical results for arbitrary $\alpha =1.5$, $w = 0.5$ and $G = 0.01$ values and different dimensions are given in Fig.\,\ref{fig-2}. Here we set $w$ in the interval $-0.5 <w \le -1$ so that it consistent with observational data. As it can be seen from that cosmological constant $\Lambda$ slowly decreases in all dimensions, however, they decay to different constant values. For $m = 2$ time evolution behavior of the cosmological constant $\Lambda$ consistent behavior with standard cosmology data for accelerated era. In higher dimensions, decaying of this parameter like the curve of $m=2$ as expected.   
\begin{figure} [ht!]
	\centering
	\includegraphics[width=6cm]{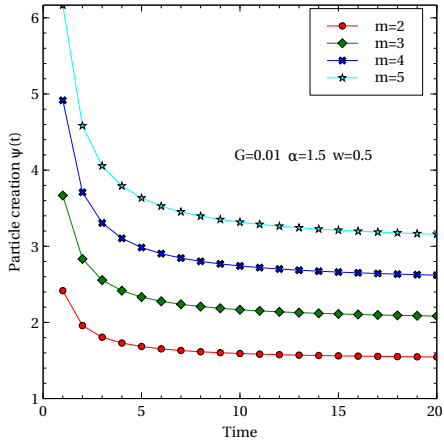}
	\caption{Particle creation versus time for $\alpha =1.5$, $w = 0.5$ and $G = 0.01$.}
	\label{fig-3}
\end{figure}

\begin{figure} [ht!]
	\centering
	\includegraphics[width=6cm]{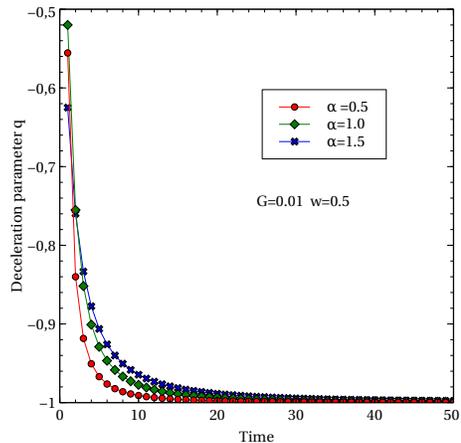}
	\caption{Deceleration parameter versus time for $w = 0.5$ and $G = 0.01$ and different $\alpha$ values.}
	\label{fig-4}
\end{figure}

To investigate the evolution of the particle creation with time for the different dimensions and arbitrary $w = 0.5$ and $\alpha = 1.5$ values, we solved Eq.\,(\ref{N-x29}) numerically. Results for different dimensions are given in Fig.\,\ref{fig-3}.
As it can be seen from that particle creation $\psi$ slowly decreases to a constant value in all dimensions. As well in Fig.\,\ref{fig-2}, time evolution behavior of the particle creation $\psi$ consistent with predictions of the standard cosmology data. The variation of deceleration parameter in compliance with time in the range of $0 < \alpha <2$ is given in Fig.\,\ref{fig-4}. Time evolution behavior in Fig.\,\ref{fig-4} consistent with standard cosmology data for $m = 2$  \cite{Pradhan2006,Pradhan2007}.
\begin{figure} [ht!]
	\centering
	\includegraphics[width=6cm]{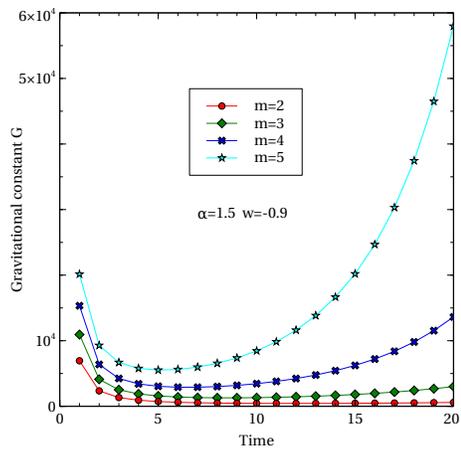}
	\caption{Gravitational constant versus time for different $-0.7 \le w \le -0.9$ values.}
	\label{fig-5}
\end{figure}
\begin{figure} [ht!]
	\centering
	\includegraphics[width=6cm]{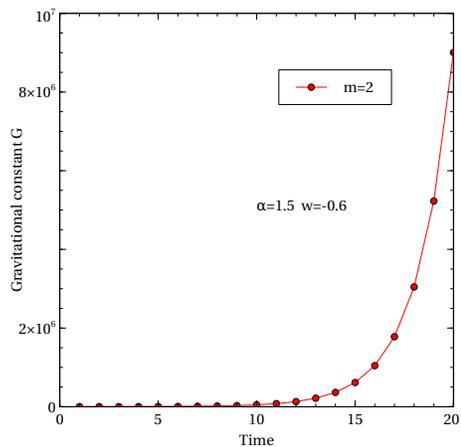}
	\caption{Gravitational constant versus time for $-0.6 \le w$ value.}
	\label{fig-6}
\end{figure}

In the case of no particle creation, using Eq.\,(\ref{N-x34}), we analyze the changing of the gravitational constant versus time in Fig.\,\ref{fig-5} for $\alpha=1.5$ and $w=-0.9$. As it can be seen from figure that for $m\ge 3$, the gravitational constant decreases for small time, however, it increases after a certain value of time while for $m=2$ gravitational parameter decreases to a constant value. We note that increasing of gravitational constant for $m\ge 3$ is very interesting result. We see that the behavior of the gravitational constant is depends on $w$, sensitively. Indeed, gravitational constant for $m\ge 3$ is a decreasing-increasing function in the interval $-0.7 \le w < -1.0$ in Fig.\,\ref{fig-5}, while it is only increasing function for all dimensions in the interval $-0.7 < w$ in Fig.\,\ref{fig-6} .
\begin{figure} [ht!]
	\centering
	\includegraphics[width=6cm]{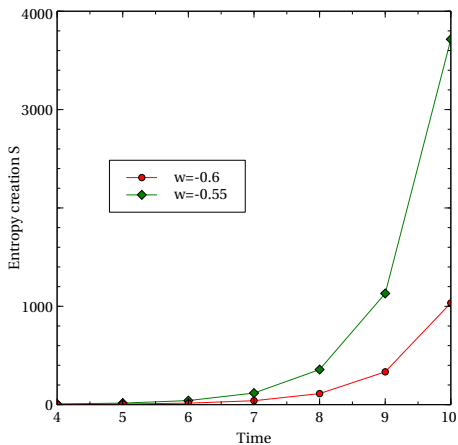}
	\caption{Entropy production versus time for $-0.6 \le w$, $m=2$, $\alpha=1.5$  and $S_{0}=1$.}
	\label{fig-7}
\end{figure}

To obtain time dependence of the entropy, we numerically solved Eqs.\,(\ref{N-e1}). Results for arbitrary different $w$ parameters is given in Fig.\,\ref{fig-7}. This figure clearly shows that entropy of the universe increases with time. This result is consistent with second law of thermodynamics. Particle creation in the universe leads to an irreversible process in which entropy always increases. Obtained results are consistent with observational data \cite{xKay2004}.

\subsection{Model II} 

The accelerating expansion of the universe at present and also its decelerating expansion in the early time form a fundamental fact which leads the motivation to take deceleration parameter. Eq.\,(\ref{N-x18}) may be rewritten as
\begin{eqnarray} \label{N-x36}
\frac{\dot{a}}{a} + q \frac{\dot{a}}{a^{2}} = 0
\end{eqnarray}
In order to solve this equation, we assume $q=q(t)=q(a(t))$. Following Chand et al. \cite{Chand2016}, the second ansatz for the scale factor can be taken as;   
\begin{eqnarray} \label{N-x37}
a(t) = \sqrt{\sinh(kt)} \ .
\end{eqnarray}
This scale factor is used to study evolution of the Bianchi Type-1 cosmology by Chawla et al. in 2012 \cite{Chawlw2012}. In addition, Pradhan \textit{et al.} used this scale factor for FRW and Bianchi models \cite{Pradhan2014b,xPradhan2012,xTiwari2017,xBiski2016}. When we use the scale factor given by Eq.~(\ref{N-x37}) in the energy density and cosmological constant in the general form that we have obtained from Eqs.\,(\ref{FE-x10}) and (\ref{FE-x11}) as
\begin{eqnarray} \label{N-x38}
\rho(t) = \frac{mk^{2}}{16 \pi G (1+w) } \sinh^{-2}(kt)
\end{eqnarray}
and
\begin{eqnarray} \label{N-x39}
\Lambda (t) = \frac{k^{2}m(m+1)}{8} \coth^{2}(kt) - \frac{km}{2}C \sinh^{-2}(kt)
\end{eqnarray}
where $C=\frac{k}{(1+w)}$. 
In addition, we have found the particle creation function like Eq.~(\ref{N-x29}) for Model II using Eq.~(\ref{N-x37}) as follow,
\begin{eqnarray} \label{N-x40}
\psi(t) = C \left( k^{2} \sinh^{-2}(kt) + \frac{(m+1)(1+w)}{2} \right) \coth(kt)
\end{eqnarray}
It is remarkable to add here that  following Model I, DP is obtained for Model II as
\begin{eqnarray} \label{N-x41}
q = -1 + 2 \sec^{2} (kt) \ .
\end{eqnarray}
\begin{figure} [h!]
	\centering
	\includegraphics[width=6cmm]{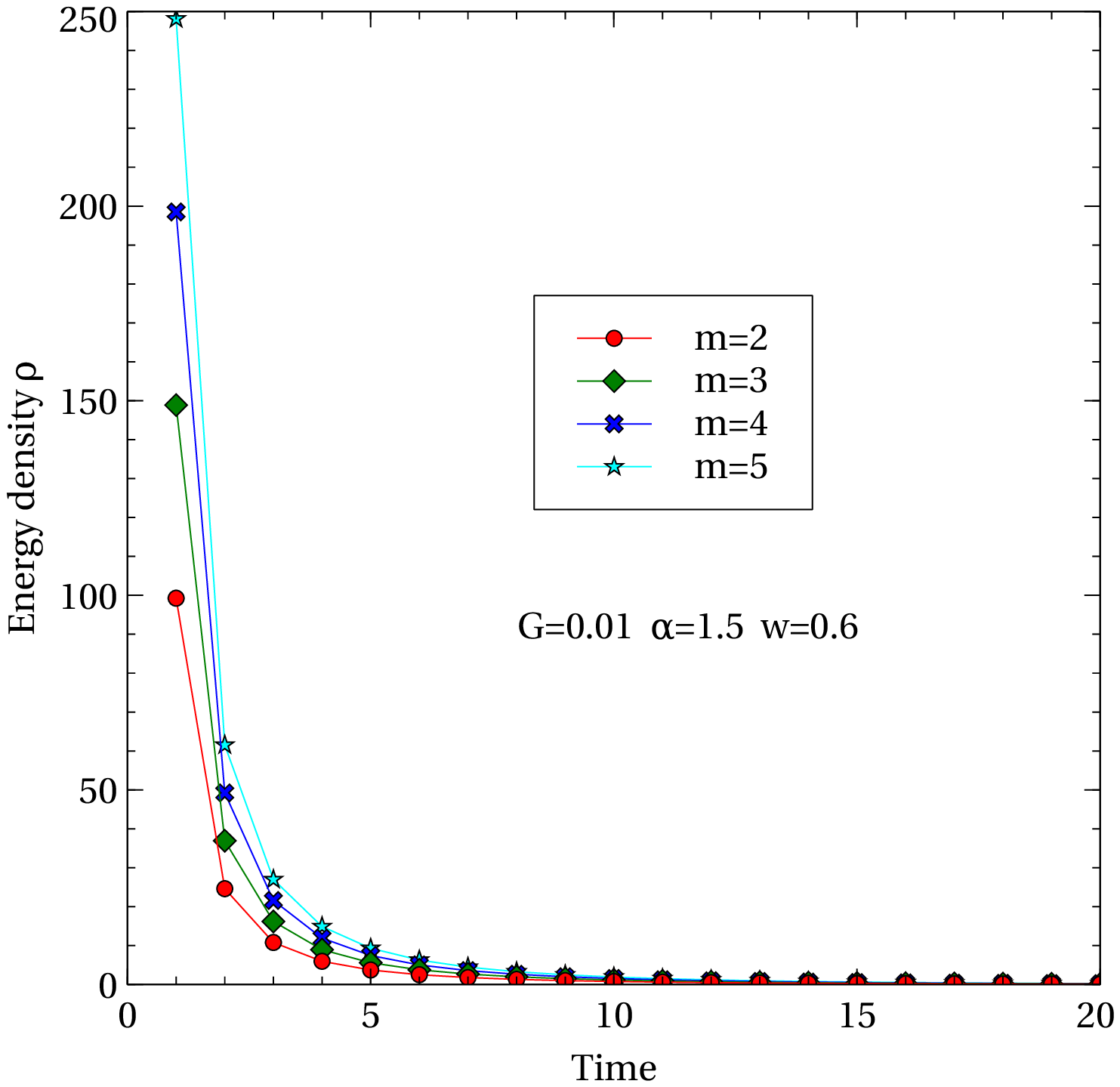}
	\caption{Energy density versus time for $\alpha=\frac{3}{2}$, $w=-0.9$ and $G=0.01$.}
	\label{fig-8}
\end{figure}
\begin{figure} [h!]
	\centering
	\includegraphics[width=6cm]{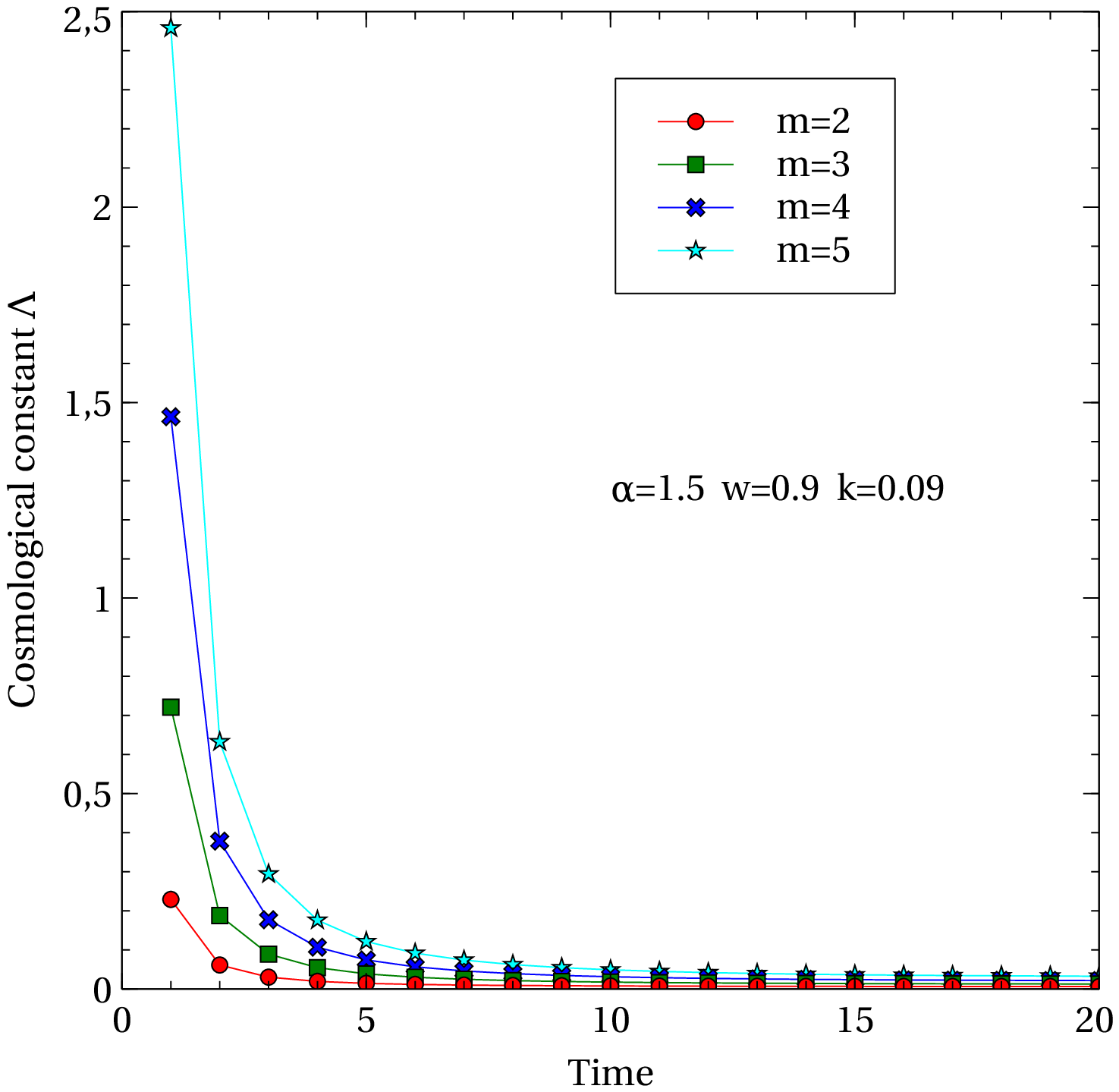}
	\caption{Cosmological constant versus time for $\alpha =1.5$, $w = 0.5$ and $G = 0.01$.}
	\label{fig-9}
\end{figure}
\begin{figure} [h!]
	\centering
	\includegraphics[width=6cm]{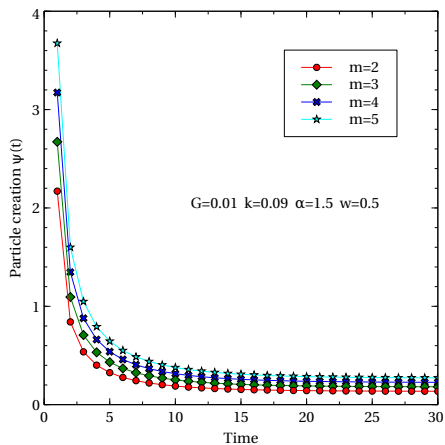}
	\caption{Particle creation versus time for $\alpha =1.5$, $w = 0.5$ and $G = 0.01$.}
	\label{fig-10}
\end{figure}

The time dependence behavior of the energy density $\rho$ and cosmological constant $\Lambda$ are given in Figs.\,\ref{fig-8} and \ref{fig-9}. For both figures, the number of dimensions  are increasing from $m = 2$ to 5. 
In Fig.\,\ref{fig-8}, we observe that $\rho(t)$ is decreasing as the time goes on and approaches to zero with persistent rate. It is clearly that energy density was very high at the beginning of the time and approaches to zero as time passes. However, if we look to Model I we can clearly see that particle creation rate differs more then Model II with the contributions of higher dimensions.

\section{Conclusion}

In the present work, the mechanism of particle creation using varying gravitational constant $G$, cosmological constant $\Lambda$ and deceleration parameter $q$ has been studied for $d$-dimensional FRW space-time to explain early deceleration and present acceleration phases. In order to investigate the dynamics of two phases, we have considered two different scale factor of the form  and  which are general form of power law expansions. First, we modified the time-varying Einstein field equations and introduced the general formulation of time-varying particle formation and entropy production mechanisms. Then,  we investigate time dependence of several cosmological constant and quantities such as particle creation function $\psi$, entropy $S$ and energy density $\rho$ etc. for two model.

In model I, energy density, cosmological constant and particle creation function behaviors were determined by the number of dimensions and time. Similar results were obtained with \cite{Singh2012,Chand2016,Amirhashchi2014} for $m = 2$. In particular, the gravitational constant for $m = 2$ and $w = -0.9$ showed a decreasing behavior in time, while $w = -0.6$ showed an exponential increase in time. In addition, an increment in time for different $w$ values of entropy, similar results were obtained with \cite{xKay2004}. When we compare equations (28) and (40), it is a very important result that the cosmological constant exhibits an exponentially decreasing behavior independently of the formation of particle creation mechanism.

In model II, a more general form of scale factor was used than model I, and energy density, cosmological constant and particle creation function behaviors were determined by the number of dimensions and time.  However, similar results were obtained with model I, as with scale factors in which we could obtain the time-dependent deceleration parameter. 

It is common practice to consider the constant deceleration parameter in the literature. Now, for a universe that has slowed down in the past and accelerated today, the DP shows a signature flip,  as discussed earlier in Section 4. Therefore, accepting the DP as variable over time can be physically verified. It is observed that Model I and II are also good harmony with observation. However, when the particle creation mechanisms shown in Fig.\,\ref{fig-3} and \ref{fig-10} are compared, it has been found that Model I shows additional contributions in higher dimensions than in Model II. 

In our subsequent study, we will analyze effects of the particle creation mechanism for Gauss-Bonnet model with high dimensional like $f(R,T)$ gravity \cite{Singh2016}.

%\section*{Acknowledgments}

\end{document}